\documentstyle[psfig]{l-aa}

\newcommand{\etal}{{\em et al.~}}
\newcommand{\ce}{Cepheid~}
\newcommand{\cs}{Cepheids~}
\newcommand{\kms}{km.s$^{-1}~$}

\begin{document}
\thesaurus{08.06.3,08.05.2,12.04.1,12.07.01}
\title{ The effect of metallicity on the Cepheid distance scale and its implications for 
the Hubble constant ($H_0$) determination
}

\author { J.P. Beaulieu\inst{1}, D.D. Sasselov$^{2,1}$, C. Renault\inst{3}, P. Grison\inst{1}, 
R.Ferlet\inst{1}, A. Vidal-Madjar\inst{1}, E. Maurice\inst{4}, L. Pr\'evot\inst{4}, 
E. Aubourg$^{3}$, P. Bareyre\inst{3}, S. Brehin\inst{3}, C. Coutures\inst{3}, 
N. Delabrouille\inst{3}, J. de Kat\inst{3}, M. Gros\inst{3}, B. Laurent\inst{3}, 
M. Lachi\`eze-Rey\inst{3}, E. Lesquoy\inst{3}, C. Magneville\inst{3}, 
A. Milsztajn\inst{3}, L. Moscoso\inst{3}, F. Queinnec\inst{3}, 
J. Rich\inst{3} ,M. Spiro\inst{3}, L. Vigroux\inst{3}, S. Zylberajch\inst{3}, 
R. Ansari\inst{5}, F. Cavalier\inst{5}, M. Moniez\inst{5},
C. Gry\inst{6}, J. Guibert\inst{7}, O. Moreau\inst{7} and F. Tajhmady\inst{7}\\
(The EROS collaboration).}
\institute{ 
Institut d'Astrophysique de Paris, CNRS, 98bis Boulevard. Arago, 75014 Paris, France
\and
Harvard-Smithsonian Center for Astrophysics, 60 Garden St., Cambridge, MA 02138, USA. 
\and
CEA, DSM/DAPNIA, Centre d'\'etudes de Saclay, 91191 Gif-sur-Yvette, France.
\and
Observatoire de Marseille, 2 place Le Verrier, 13238 Marseille 03, France.
\and
Laboratoire de l'Acc\'el\'erateur Lin\'eaire IN2P3, Centre d'Orsay, 91405 Orsay, France.
\and
Laboratoire d'Astronomie Spatiale CNRS, Travers\'ee du siphon, les trois lucs, 13120 Marseille, France.
\and
Centre d'Analyse des Images de l'Institut National des Sciences de l'Univers, 
CNRS Observatoire de Paris, 61 Avenue de l'Observatoire, 75014 Paris, France. }
\offprints{JP. Beaulieu}

\date{Received;accepted}
\maketitle
\markboth{J.P.\ Beaulieu et al.: Metallicity effect on the Cepheid distace scale}{J.P.\ Beaulieu et al.: Metallicity effect on the Cepheid distace scale}

\begin{abstract}
Recent HST determinations of the expansion's rate of the Universe (the Hubble constant, $H_0$) 
 assumed that the  Cepheid Period-Luminosity relation at V and I are independent of metallicity 
(Freedman, et al., 1996, Saha et al., 1996, Tanvir et al., 1995).
The three groups obtain different vales for $H_0$. We note that most of this discrepancy stems from the asumption
(by both groups) that the Period-Luminosity relation is 
independent of metallicity. We come to this conclusion
as a result of our study of the Period-Luminosity relation   of  481 
Cepheids with 3 millions two colour measurements in the Large Magellanic
Cloud and the Small Magellanic Cloud obtained as a by-product of the EROS microlensing survey.
We find that the derived interstellar absorption corrections 
are particularly sensitive to the metallicity and when our result is
applied to recent estimates based on HST Cepheids observations it
makes the low-$H_0$ values higher and the high-$H_0$ value lower, bringing those
discrepant estimates into agrement around $H_0 \approx 70$ kms$^{-1}$Mpc$^{-1}$. 
\\

\end{abstract}

\section {Introduction}

The value of the Hubble constant $H_0$ fixes the fundamental physical 
scale and time scale of the Universe, therefore by inference its age.  
Dichotomous results are obtained in discordant ranges, even allowing for their 
internal error estimates. The results are dependent upon the choice of the cosmological 
distance ladder used (Fukugita et al., 1993, Freedman et al., 1996, Saha et al., 1996). However, both 
 the longer distance scale ($H_0 \approx 50$ kms$^{-1}$Mpc$^{-1}$), 
and the shorter distance scale ($H_0 \approx 80$ kms$^{-1}$Mpc$^{-1}$) rely strongly on 
the Cepheid period-luminosity relation calibrated in the Large Magellanic Cloud (LMC),
and assumed to be independent of metallicity. 
The Cepheid Period-Luminosity (PL) relation is widely considered to be the most accurate
way available to measure extragalactic distances on small scale. The Cepheids are relatively 
young, bright periodic variable stars which pulsational periods are strongly 
correlated with their luminosity. Their distance is therefore determined via a comparison 
to a calibrated PL relation.
Cepheids are observed in distant galaxies, up to the Virgo cluster, in order
to calibrate secondary distance indicators that  connect to the Hubble flow.

The theory (Stothers, 1988, Stift 1990, Chiosi et al., 1993, Stift 1995) 
predicts a small abundance ( helium fraction and heavy elements) effect on the  PL zero point due to : 
(i)\phantom{ii}  theory of stellar pulsation  through the dependence of period on mass and radius.
(ii)\phantom{i} theory of stellar evolution  through the mass luminosity relation.
(iii) theory of stellar atmosphere through the relation between effective temperature,
absolute magnitude in bandpasses and bolometric correction.
Disentangling the effects of metallicity and reddening has a venerable history (Feast 1991).
Various observational studies since the 70s, have looked for and found a color shift between LMC and SMC
Cepheids (e.g. Gascoigne 1974, Martin et al. 1979). Caldwell \& Coulson (1986) and more recently
Laney \& Stobie (1994) used theoretical and empirical relations, as well as individual reddenings
from color-color diagrams to derive PL relations adjusted for abundance differences. 
Their approach is different from ours (and the technique used by 
the $HST$ teams); a full comparison will be model dependent (flux 
redistribution models) and we have not attempted to do that here. A limited
comparison to the VJHK Table 6 of Laney \& Stobie (1994) shows that we confirm
the sense of their results.

An empirical test (Freedman and Madore, 1990, Gould, 1994)
in three fields of M31 with 36 Cepheids and 152 BVRI measurements have led to ambiguous 
results. 
As part of the HST key project on extragalactic distance scale observations, a further
check is planned by observations of two fields in M101.
Currently, all HST studies assume that the Cepheid PL relation has no metallicity dependence
(Freedman et al., 1994a, Freedman et al., 1994b, Sandage et al., 1994, Tanvir et al., 1995,
Kennicutt et al., 1995).

\section { EROS observations.}

We used a new data set of 481 Cepheids and 3 millions measurements in two colours
obtained in the LMC and the SMC to derive the dependence of the optical period 
luminosity relations on metallicity. Observations were obtained as a by-product
of the EROS (Exp\'erience de Recherche d'Objets Sombres, Aubourg et al., 1993) microlensing survey. 

Observations have been obtained from ESO La Silla using a 0.4m f/10 reflecting telescope
and a $2 \times 8$ mosaic of CCDs (Arnaud et al., 1994ab). During the $\sim 100$ nights 
of 1991-1992 campaign (Beaulieu et al., 1995, Grison et al., 1995), 
about 2000 images have been obtained in the bar of the LMC in a blue and a red bandpass 
($B_E$ and $R_E$), whereas about  6000 images have been obtained in a crowded field of 
the SMC with a near pair of very similar filters ($B_{E2}$ and $R_{E2}$) during the 
$\sim 200$ nights of 1994-1995 season (Beaulieu et al., 1996). 
An accurate and reliable photometric transformation have been constructed between  the two systems.
Hence we present a differential analysis that relies on no external zero-point and thus 
completed within the ($B_E$ and $R_E$) system.
Since we apply our result to HST $V$, $I$ observations, we make use of the fact that
the net transmission of the $B_E$ band after convolution with the CCD reponse is much closer 
to Johnson $V$ than to Johnson $B$. The transformation is similar to that between the
broader $HST$ F555W filter and $V$. The $R_E$ band is between Cousin $R$ and $I$.
The $B_E-R_E$ color transforms well to $V-I$ : $V-I = 1.02(B_E-R_E),~\sigma=0.02$mag.

The two data sets were searched for Cepheids with Fourier analysis (Grison 1994) and phase dispersion 
minimisation techniques (Schwarzenberg-Czerny 1989).
Stars affected by blending effects have been excluded of the sample. 
The high quality, excellent phase coverage light curves were Fourier analysed to separate 
between the classical Cepheids that pulsate in the fundamental mode, and the so called
s-Cepheids that pulsate in the first overtone, and therefore follow a different period 
luminosity relation. In the LMC we keep 51 fundamental pulsators and 27 first overtone 
pulsators, and 264 fundamental pulsators and 141 first overtone pulsators in the SMC.
So, we have two complete samples of Cepheids that fill densely the 
period-magnitude-color [PLC] space, with known difference
in metallicity $\Delta [Fe/H]_{LMC-SMC} = 0.35$ (Spite and Spite, 1991, Luck and Lambert 1992). 

\section {The method}

We compare the two samples in the 
PLC manifold to derive 2 independant sources of difference
$-$ distance and extinction, and to search for a third one $-$ metallicity.
The method will be applied independently to both fundamental and first 
overtone pulsators, and is a $\chi^2$ minimization fit to a model. Details
of the method are presented elsewhere (Sasselov et al., 1996).
The model is built on the basis
of the technique used for determining Cepheid distances to galaxies 
by the $HST$ Key Project (Freedman and Madore, 1990). It is as follows: 
the LMC PL relation is used to slide the observed PL relation against it;
the Galactic extinction law is applied in deriving the reddening 
(implicitly assuming
$no$ difference between the ensemble colors of the Cepheids in LMC 
and the observed
galaxy); a true distance modulus is derived by adopting a LMC distance 
and correcting
for extinction (using the reddening derived above with adopted LMC 
reddening E(B-V)=0.10).
Other assumptions in the technique (and our model) are: 
constant PL slope with metallicity
(confirmed by our observations) and no depth dispersion in the LMC Cepheid
sample (which is only from the LMC bar).
In the application we use as constrains the independently derived foreground
reddening of LMC (0.06) and SMC (0.05), and the line of nodes for SMC from
Caldwell \& Coulson (1986) to derive the EROS SMC sample depth dispersion.
We fit for different fixed values of the extinction parameter $R_V$.
The model describes the case considered by Stothers (1988) in his equation (26).

We model the $B_E$, $R_E$ data in the PLC space taking into 
account the high degree of correlation between the measurements following
the formalism introduced by Gould (1994). Unlike Gould, we solve for a
wavelength dependance of a metallicity effect.
 Our model has 12 parameters, which are : 
linear fits of the corelations in the PLC projections 
(slopes $\beta_i$,$b_i$ and the zero points $\alpha_i$,$a_i$),
the distance difference $\gamma_1$,  
the relative reddening difference $\gamma_2$, 
and the metallicity terms $\gamma_3^i$ ($i$=1,2 are the two passbands,
$B_E$ and $R_E$). The model parameters are estimated by minimizing the
mean magnitude residuals, $X_{i,p}$, where $Q_{i,p}$ is the observed
mean magnitude in the $i$ band for a Cepheid with period $P_p$.
For example, the residuals for the SMC sample are given as:

\begin{equation}
X_{i,p}^3=Q_{i,p}-(\alpha_i+\beta_ilogP_p+\gamma_1+\gamma_2R_i+\gamma_3^i),
\end{equation}
$$
X_{i,p}^4=Q_{i,p}-[a_i+b_i(Q_{1,p}-Q_{2,p}+\gamma_2(R_2-R_1)~~~~~~~~~~~~~~~~
$$
\begin{equation}
~~~~~~~~~~~~~~~+\gamma_3^1-\gamma_3^2)+\gamma_1+\gamma_2R_i+\gamma_3^i],
\end{equation}
 with $p=1,...,N(n)$ being the number of Cepheids.
The form of the covariance matrices of the data, the $\chi^2$ minimization and 
the iteration procedure are described in Sasselov et al. (1996).

\section { The results}

We applied the global fitting procedure to the data-set of fundamental mode and
first overtone mode pulsators independently, and we obtained the same results.\\

The two set of PL relation give, before correction for metallicity and reddening, 
the same distance modulus difference of $\delta ({LMC-SMC}) = 0.89 \pm 0.05$mag.
The other model parameters have the following values for the fundamental pulsators : 
$ \alpha_i, \beta_i = 17.61 \pm 0.035, -2.72 \pm 0.07$; $17.74 \pm 0.029,-2.95 \pm 0.06;$
$a_i, b_i = 14.49 \pm 0.13, -11.78 \pm 0.94$; $14.36 \pm 0.12 -12.77 \pm 0.90$.
$\gamma_1 = 0.62 \pm 0.04, \gamma_2 = 0.01 \pm 0.01$, $\gamma_3^1 = 0.06 \pm 0.01$,
$\gamma_3^2 = -0.01 \pm 0.01$.
We adopted the same extinction law with $R_V = 3.3$ for the SMC.
The correction due to the  metallicity dependence of the inferred distance modulus by the described 
technique is then $\delta \mu= \Delta \mu_{true}-\Delta \mu_{infered}=-0.139^{+0.04}_{-0.06}$mag.

The sources of error are the photometric transformation, between the two data sets 
(0.06 mag), the scatter in the PL relation (0.04 mag) and the uncertainties in 
the reddening and depth determination in the SMC (0.03 mag).

An alternative to the metallicity effect could be unusual SMC extinction with
a value of $R_V \ge 5$ and 
$E(B-V)\leq 0.04$. We consider this alternative to be very unlikely, because
it requires that the extinction be less than the foreground to SMC and an
unusually high $R_V$ value unjustified by any other evidence.
On the other hand, with $R_V=3.3$, our mean reddening value is in good 
agreement with a number of other independent estimates (Bessell 1991). 
 
We derive the following metallicity dependence of reddening
corrected distance moduli of Cepheids:
$$\delta {\mu}=(0.44_{-0.2}^{+0.1}) ~log\frac{Z}{Z_{LMC}},$$ where Z is the abundance of 
heavy elements (by mass) in the studied \cs and $Z_{LMC}$=0.0085. 
The metallicity dependence is valid in the spectral region covered by the EROS 
filters,  but can be applied to HST (V,I) work for the reasons presented in Sect.2
above. The form of the metallicity dependence used here is based on theoretical
considerations and is a good assumption in the suggested range of application
of $0.001\leq Z \leq 0.02$ (Chiosi \etal 1993).

Despite its relative weakness, this metallicity effect is significant to extragalactic distance 
measurements. This is especially true when the color shift due to  metallicity 
$\bf \gamma_3^1-\gamma_3^2$ is  interpretated as reddening in the determination of the 
true distance modulus of a target galaxy
(Freedman et al., 1994a, Freedman et al., 1994b, Sandage et al., 1994, Tanvir et al., 1995,
Kennicutt et al., 1995).\\

\section { Effect on $H_0$ determination }

 Recent efforts to determine $H_0$ from \ce distances have focused on:
(1) the Virgo cluster galaxies (Freedman et al., 1994a),
(2) the Leo~I group (Tanvir et al., 1995), containing elliptical galaxies; and
(3) the parent galaxies (Sandage et al., 1994) of supernovae of type Ia.
All these studies use the same modern technique with the LMC as a base
and all the same initial assumptions; we share them in our analysis.
Yet they result in three different values of $H_0$, with hardly overlaping
error bars.
The abundances of metals in the young populations of the host galaxies
for these studies differ and, apparently, in a systematic way. 
Therefore we propose that the metallicity dependence may be responsible
for most of this discrepancy.

With the first approach and HST $VI$ photometry of \cs in M100, they
derive $H_0=80\pm17$ \kms $Mpc^{-1}$. For the metallicity
of the \cs we adopt [Fe/H]=+0.1 (Z=0.02), derived from the abundances of

H~II regions (Zaritsky et al., 1994) and assuming a solar [O/Fe] ratio.
With the uncertainty range of our metallicity effect, this leads to
$H_0=76-70$ \kms $Mpc^{-1}$.

With the second approach and HST $VI$ photometry of \cs in M96 (Tanvir et al., 1995),
they derive $H_0=69\pm8$ \kms $Mpc^{-1}$. 
Oey \& Kennicutt (1991) give an abundance estimate for M96 of [Fe/H]=-0.02.
This implies a small change in the \ce distance to M96 and the value of $H_0$, $H_0=64-66$ \kms $Mpc^{-1}$.

With the third approach and HST $VI$ photometry of \cs in IC4182 and NGC5253 (Sandage et al., 1994),
they derive $H_0=55\pm8$ \kms $Mpc^{-1}$. Abundances in
NGC5253 have been measured and discussed (Pagel et al., 1992);  discussion
of abundances in IC4182 has been also given by Saha et al., (1996) $-$ we adopt
[Fe/H]=-1.3 and caution on the large uncertainties. With this metallicity
$H_0=59-68$ \kms $Mpc^{-1}$. Other systematic effects in the SN~Ia method (Riess et al., 1995),
bring the value of $H_0$ even further up; to a value also favored by
new theoretical SN models (H\"oflich and Khokhlov 1996). 
The new value of $H_0=58\pm4$ \kms $Mpc^{-1}$ (Sandage et al., 1996)
includes two SNe~Ia in spiral galaxies, which Cepheid metallicity is most
likely similar to that of the LMC. Note, however, that the two SN (1981B
and 1990N) are by 0.2$-$0.4 $mags$ dimmer than the three SNs in IC4182 and
NGC5253. Therefore, the result we obtained above by accounting for the effect
of metallicity on the Cepheid distances for each galaxy, remains virtually
unchanged.

The metallicity dependence we found from the LMC/SMC analysis brings all
the derivations of $H_0$ to good agreement. It is summarized in Fig. 2.

\begin{figure}
\includegraphics{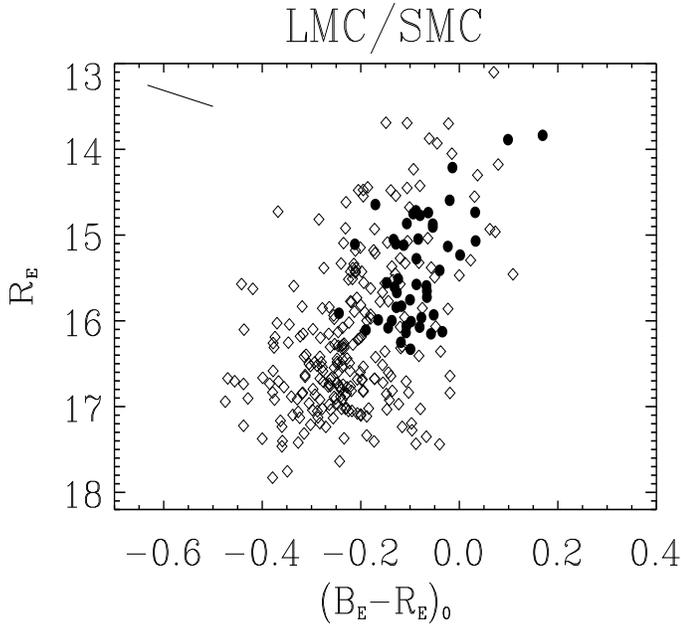}
\vspace*{3.2in}
\caption{The colour magnitude diagram for the fundamental mode Cepheid
pulsators in the SMC (diamond) and LMC (filled diamond) in the 
$B_E-R_E$ colour system.
The difference vector represents 
the reddening line in this plane. The color difference between the two
samples of Cepheids is due to their different metallicity. The SMC Cepheids
are bluer and fainter at a given pulsation period.}
\end{figure}

\begin{figure}
\includegraphics{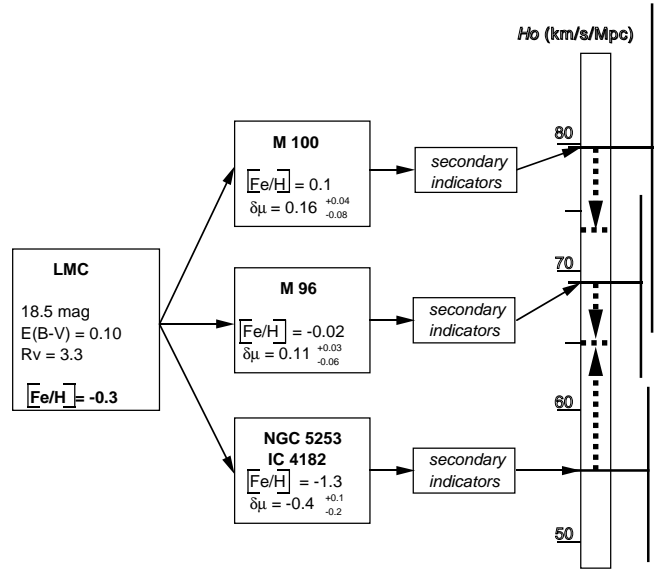}
\vspace*{2.8in}
\caption{Metallicity effect on the Cepheid extragalactic distance scale and
its influence on HST observations of distant galaxies : all these $H_0$
determination are brought into good agreement.
the quoted uncertainty in $\delta mu$ comes from our uncertainty of the 
metallicity effect. We do not include uncertainty in [Fe/H].}
\end{figure}

\begin{acknowledgements}
We are grateful for the support given to our project by the technical staff at ESO La Silla.

\end{acknowledgements}

\end{document}